\documentclass[preprint,aps,showpacs,nofootinbib]{revtex4}

\usepackage{amssymb}
\usepackage{color}
\usepackage{epsfig}
\definecolor{White}{rgb}{1,1,1}
\definecolor{Red}{rgb}{1,0.1,0}
\definecolor{LightYellow}{rgb}{1,1,.875}
\definecolor{SteelBlue}{rgb}{.273,.508,.703}
\definecolor{navy}{rgb}{0,0,.5}
\definecolor{LightCyan}{rgb}{.875,1,1}
\definecolor{DarkRed}{rgb}{.543,0,0}
\definecolor{HotPink}{rgb}{1,.41,.70}
\definecolor{ForestGreen}{rgb}{.13,.54,.13}
\definecolor{OliveDrab}{rgb}{.42,.55,.14}
\definecolor{MediumBlue}{rgb}{0,0,.80}
\definecolor{RoyalBlue}{rgb}{.25,.41,.88}
\definecolor{DeepSkyBlue}{rgb}{0,.746,1}
\definecolor{Brown}{rgb}{0.545,0.271,0.074}

\def\bea{\begin{eqnarray}}
\def\eea{\end{eqnarray}}
\def\bec{\begin{center}}
\def\ec{\end{center}}

\def\beq{\begin{equation}}
\def\eeq{\end{equation}}
\def\haf{\frac{1}{2}}

\begin{document}

\begin{flushright}
\today
\end{flushright}

\title{\Large Singlet Fermionic Dark Matter explains DAMA signal}

\author{Yeong Gyun Kim and Seodong Shin}

\vskip 2cm

\affiliation{{\it Department of Physics, KAIST, Daejeon 305-701, Korea}}

\begin{abstract}
It has been suggested that, considering channeling effect, the order of a few GeV dark matters 
which are elastically scattered from detector nuclei 
might be plausible candidates 
reconciling the DAMA annual modulation signal 
with the results of other null experiments. 
We show that Singlet Fermionic Dark Matter can be 
such a dark matter candidate, simultaneously providing
the correct thermal relic density which is consistent with the WMAP data. 
\end{abstract}

\maketitle

\section{Introduction}

We have now compelling evidences for the existence of non-baryonic dark matter (DM) in the universe,
whose mass density has been accurately measured by the Wilkinson Microwave
Anisotropy Probe (WMAP) \cite{wmap}.
The DM mass density normalized by the critical density is given by, 
in the $3\sigma$ range, 
\begin{eqnarray}
0.091 < \Omega_{\rm CDM} h^2 < 0.129, 
\label{wmap}
\end{eqnarray}
where $h \approx 0.7$ is the scaled Hubble constant in the units of 100 (km/sec)/Mpc. 

Recently, the DAMA collaboration has reported the observation of 
an annual modulation \cite{am} in nuclear recoil rate of NaI(Tl) detectors \cite{dama}. 
The DAMA signal is consistent with elastic, spin independent DM scattering 
from target nuclei in the detectors. 
The conventional signal region corresponds to the DM mass and 
the scattering cross section 
$(m_{DM}, \sigma^{SI}_{p}) \sim (30-200\, {\rm GeV}, 10^{-5}\, {\rm pb})$,
which is now excluded by other DM search experiments 
such as XENON10 \cite{xenon10} and CDMS (Ge) \cite{cdms}.

It has been shown, however, that considering effect of channeling \cite{channeling1, zurek} 
in the NaI crystal scintillators of DAMA, 
the spin-independent elastic scattering of DM with nuclei can accommodate the DAMA signal 
with the results of other null experiments. 
The corresponding region of the cross section, 
which might be compatible with all experiments 
(even without considering dark stream \cite{darkstream}), 
is given by \cite{zurek}
\begin{eqnarray}
3 \times 10^{-41} {\rm cm^2} \lesssim \sigma_p^{SI} \lesssim 
5 \times 10^{-39} {\rm cm^2},
\label{sigma}
\end{eqnarray}
with a DM mass in the range of
\begin{eqnarray}
3~ {\rm GeV} \lesssim m_{DM} \lesssim 8~ {\rm GeV}.
\label{mass}
\end{eqnarray}

Various models have been studied to accommodate the DAMA signal region 
with the parameters of (\ref{sigma}) and (\ref{mass}), 
which include mirror dark matter \cite{foot}, 
WIMPless dark matter \cite{wimpless},
light neutralino in the minimal supersymmetric standard model (MSSM) \cite{bottino} 
and right-handed sneutrino dark matter from Next-to-MSSM (NMSSM) \cite{cerdeno}. 
Adding to those models, there are other good approaches explaining 
the DAMA modulation signals as massive WIMPs (heavier than 100 GeV) 
free from null experimental bounds, which focus on inelastic dark matter 
with non-standard halo models of DM velocity distribution or 
various galactic escape velocities \cite{iDM}. 

In this work, we would like to direct a spotlight on 
a singlet fermionic dark matter \cite{kim2,kim1} elastically scattering 
from target nuclei with (\ref{sigma}) and (\ref{mass}) to explain 
the DAMA signal reconciling other null experiments, 
which can also simultaneously satisfy 
the measured DM mass density $(\ref{wmap})$. 
Our paper is organized as following. 
In section \ref{sec:Model}, we briefly review the model of singlet 
fermionic dark matter. For other recent studies of
singlet . Model parameter regions, which explain the DAMA signal 
in the low mass region with the correct thermal relic density, 
are investigated in section \ref{sec:NA}. 
Section \ref{sec:conclusion} is the conclusion.

\section{Singlet Fermionic Dark Matter Model}
\label{sec:Model}

A standard model (SM) gauge singlet sector, aka "hidden sector", is introduced, 
which consists of a real scalar field $S$ and a Dirac fermion field $\psi$
\cite{kim2}
\footnote{For other recent studies of hidden dark matter, see Refs. \cite{others}.}. 
The singlet scalar $S$ couples to the SM particles
only through the interactions with the SM Higgs boson. 
The interaction of the singlet DM fermion $\psi$ with the SM particles 
comes through the interaction of $\psi$ with the singlet scalar $S$ and 
the mixing of the scalar $S$ with the SM Higgs. 

The model Lagrangian is written as
\begin{eqnarray}
\mathcal{L} = \mathcal{L}_{SM} + \mathcal{L}_{hid} + \mathcal{L}_{int},
\end{eqnarray}
where $\mathcal{L}_{SM}$ stands for the SM Lagrangian and the hidden sector Lagrangian is given by
\begin{eqnarray}
\mathcal{L}_{hid} = \mathcal{L}_{S} + \mathcal{L}_{\psi}
                       -g_S \bar{\psi}\psi S ,
\end{eqnarray}
with
\begin{eqnarray}
\mathcal{L}_{S} &=&
\haf \left(\partial_{\mu} S\right)\left(\partial^{\mu} S\right)
      -\frac{m_0^2}{2} S^2 -\frac{\lambda_3}{3!}S^3-\frac{\lambda_4}{4!}S^4 ,
\label{eq:ps0}
\\
\mathcal{L}_{\psi} &=& \bar{\psi}\left(i\partial\!\!\!/
                - m_{\psi_0}\right)\psi. \label{eq:ps}
\end{eqnarray}
The interaction terms between the singlet scalar $S$ and the SM Higgs $H$ are given by
\begin{eqnarray}
\mathcal{L}_{int} = -\lambda_1 H^{\dagger} H S
                   - \lambda_2 H^{\dagger} H S^2.  \label{eq:int}
\end{eqnarray}

The scalar potential given in Eq.(\ref{eq:ps0}) and (\ref{eq:int}), together with the SM Higgs potential
\begin{eqnarray}
V_{SM} = -\mu^2 H^{\dagger} H + \lambda_0 (H^{\dagger} H)^2,
\end{eqnarray}
derives vacuum expectation values, $\langle H^0 \rangle=v_0/\sqrt{2}$ and $\langle S \rangle = x_0$,
of the neutral component of the SM Higgs and the singlet scalar, respectively. 
The extremum conditions $\partial V / \partial H |_{<H^0>=v_0/\sqrt{2}}= 0$ and
$\partial V / \partial S |_{<S>=x_0} =0$, of total scalar potential $V$, lead to the following relations among the model parameters \cite{PMS}
\begin{eqnarray}
\mu^2 &=& {\lambda}_0 v_0^2 + (\lambda_1 + \lambda_2 x_0)x_0 ,
\nonumber \\
m_0^2 &=& - \frac{\lambda_3}{2}x_0 -\frac{\lambda_4}{6} x_0^2
          - \frac{\lambda_1 v_0^2}{2 x_0} - \lambda_2 v_0^2 .
\end{eqnarray}

The neutral scalar fields $h$ and $s$ defined by $H^0=(v_0+h)/\sqrt{2}$ and $S=x_0+s$ are mixed to yield the mass matrix given by
\begin{eqnarray}
\mu_{h}^2 &\equiv& \left.\frac{\partial^2 V}{\partial h^2}\right|_{h=s=0}
           = 2 {\lambda}_0 v_0^2 ,
\nonumber \\
\mu_{s}^2 &\equiv& \left.\frac{\partial^2 V}{\partial s^2}\right|_{h=s=0}
           = \frac{\lambda_3}{2} x_0 + \frac{\lambda_4}{3} x_0^2
                  - \frac{\lambda_1 v_0^2}{2 x_0} ,
\nonumber \\
\mu_{hs}^2 &\equiv& \left.\frac{\partial^2 V}
                               {\partial h \partial s}\right|_{h=s=0}
           = (\lambda_1 + 2 \lambda_2 x_0) v_0.
\end{eqnarray}
The corresponding mass eigenstates $h_1$ and $h_2$ are defined by
\begin{eqnarray}
h_1 &=& \sin \theta \ s + \cos \theta \ h ,
\nonumber \\
h_2 &=& \cos \theta \ s - \sin \theta \ h ,
\end{eqnarray}
where the mixing angle $\theta$ is given by
\begin{eqnarray}
\tan \theta = \frac{y}{1+\sqrt{1+y^2}},
\end{eqnarray}
with $ \ y \equiv  2 \mu_{hs}^2 /(\mu_h^2 - \mu_s^2)$. 
Then the two Higgs boson masses $m_1$ and $m_2$ are given by
\begin{eqnarray}
m^2_{1,2} = \frac{\mu_h^2+\mu_s^2}{2}
            \pm \frac{\mu_h^2-\mu_s^2}{2}\sqrt{1+y^2},
\end{eqnarray}
where the upper (lower) sign corresponds to $m_1(m_2)$.
From the above definition of the mixing angle $\theta$,
we get $|\cos \theta| > 1/\sqrt{2}$, hence $h_1$ is SM Higgs-like while $h_2$ is singlet-like.

In total, we have eight independent model parameters relevant for DM phenomenology. 
The six model parameters 
$\lambda_0, \lambda_1, \lambda_2, \lambda_3, \lambda_4$ and $x_0$ 
determine the masses, mixing and self-couplings of the scalar sector. 
The singlet fermion $\psi$ has mass $m_\psi = m_{\psi_0} + g_S x_0$ 
as an independent parameter of the model since $m_{\psi_0} $ is 
just a free model parameter. 
Finally, the Yukawa coupling $g_S$ measures the interaction of $\psi$ 
with singlet component of the scalar particles.

\section{Numerical Analysis}
\label{sec:NA}

\begin{figure}[ht]
\vskip 0.4cm
\begin{center}
\epsfig{figure=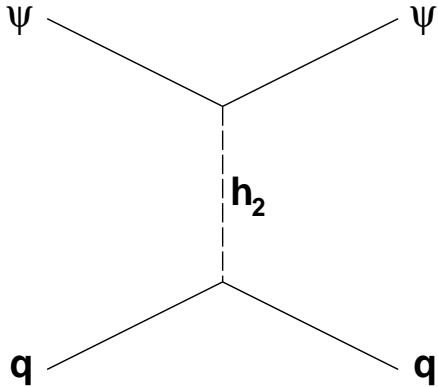,width=7cm,height=7cm}
\end{center}
\caption{\it Dominant diagram for spin-independent DM-nucleon scattering,
with a low mass of the singlet-like Higgs $h_2$.}
\label{fig:scattering}
\end{figure}

Spin-independent elastic scattering of the fermionic DM $\psi$ with nucleons arises from $t-$channel Higgs exchange diagrams, which is shown in Fig. \ref{fig:scattering}. For a light dark matter with $m_\psi \simeq 5$ GeV, 
the spin-independent cross section for DM-proton scattering is approximately given by 
\begin{eqnarray}
\sigma (\psi p \rightarrow \psi p) \sim 
0.1 \left({g_S \, {\rm cos}\theta \, {\rm sin}\theta \over v_0 \, m_{h_2}^2}\right)^2
{\rm (GeV^4)},
\end{eqnarray}
assuming $m_{h_2} << m_{h_1}$. 
With $g_S \sim O(1)$, a low Higgs mass $m_{h_2}$ is needed in order to obtain 
a desired value of $\sigma(\psi p \rightarrow \psi p) \sim 10^{-4}$ pb 
for explaining the DAMA result. On the other hand 
the scalar mixing angle $\rm sin\theta$, which determines $Z Z h_2$ coupling, 
should be small not to conflict with the LEP constraint on the Higgs mass \cite{LEP}, 
for such a low Higgs mass. 
For instance, $m_{h_2} \sim 5\, (15)$ GeV, ${\rm sin}\theta \sim 0.01\, (0.1)$ with $g_S = 1$ 
provides the DAMA favored value for the scattering cross section 
while satisfying the LEP constraint on the Higgs mass.

\begin{figure}[ht]
\vskip 0.4cm
\begin{center}
\epsfig{figure=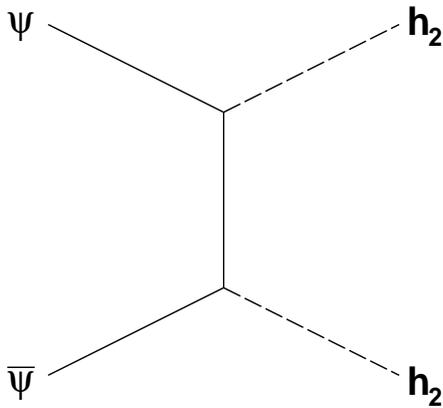,width=7cm,height=7cm}
\end{center}
\caption{\it Dominant diagram for DM pair annihilation, when $m_{\psi} \lesssim m_{h_2}$. }
\label{fig:annihilation}
\end{figure}

One concern about the light fermionic DM scenario for explaining the DAMA data 
is that it usually leads to too large 
thermal DM relic density to be compatible with the WMAP measurement (\ref{wmap}), 
if DM pair annihilation undergoes mainly through 
s-channel Higgs exchanges \cite{andreas}. 
However, when the singlet-like Higgs mass is less than or 
similar to the DM particle mass, {\it i.e.}, $m_{h_2} \lesssim m_\psi$, 
the dominant contribution for the annihilation of DM pair in the early universe 
arises from $\psi \bar\psi \rightarrow h_2 h_2$ process, 
which is illustrated in Fig. \ref{fig:annihilation}. 
In this kinematic regime, the singlet-like Higgs does not decay into the DM pair 
since it is not kinematically allowed, but it decays entirely into the SM particles. 
The DM pair annihilation to $h_2$ pair is not suppressed 
by the small mixing angle $\theta$ because $\bar\psi \psi h_2$ coupling 
is proportional to $\rm cos\theta$ rather than $\rm sin\theta$, 
thus it may provide large enough annihilation cross-section to 
have a small DM relic density compatible with the WMAP data \cite{maxim}.

As a specific numerical example, let us consider the following model parameters
\begin{eqnarray}
x_0=100\, {\rm GeV}, ~\lambda_0 = 0.12, ~ \lambda_1 = -19, ~ \lambda_2 = 0.1,
~ \lambda_3 = -314\, {\rm GeV}, ~ \lambda_4 = 3,
\end{eqnarray}
which provide the masses and mixing angle of two Higgs particles as
\begin{eqnarray}
m_{h_1} = 120.5\, {\rm GeV},~m_{h_2} = 6.7\, {\rm GeV}\, \,
{\rm and} \, \, \, \rm sin\theta = 0.017.
\end{eqnarray}
Further setting $m_\psi = 5 \,{\rm GeV}$ and $g_S=1.2$, 
we obtain the following spin-independent DM scattering cross section 
and thermal DM relic density;
\begin{eqnarray}
\sigma(\psi p \rightarrow \psi p) \sim 2 \times 10^{-4}~{\rm pb} 
~~{\rm and}~~ \Omega h^2 \sim 0.1 ,
\end{eqnarray}
which would explain the DAMA signals without conflicting other null experiments, 
and also satisfy the measured DM mass density simultaneously. 
We also note that the lifetime of the singlet-like Higgs $h_2$ 
is much shorter than one second. 
Thus it decays well before the start of the Big Bang Nucleosynthesis, 
making no cosmological problem.

\begin{figure}[ht]
\begin{center}
\epsfig{figure=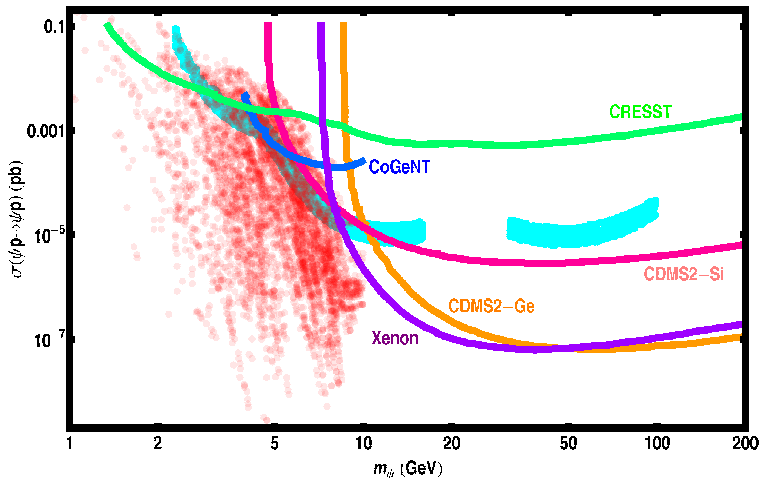,width=10cm}
\end{center}
\caption{\it Spin-independent cross section for DM-nucleon scattering as
a function of DM mass $m_{\psi}$. The red points are the predictions
for the light singlet fermionic dark matter and the cyan regions are DAMA signal regions. 
Also denoted are the upper limits from various DM search experiments.}
\label{fig:scan}
\end{figure}

Now we scan our model parameters in a certain region, 
which provides a low singlet-like Higgs mass 
(1 GeV $< m_{h_2} < 10$ GeV), $m_{h_1} \sim 120$ GeV and 
small scalar mixing angle ($|\theta| < 0.02$) 
within a low DM mass range (1 GeV $< m_{\psi} < 10$ GeV). 
Fig. \ref{fig:scan} is the result of the parameter scan, 
showing the spin-independent cross section for DM-nucleon scattering 
as a function of the singlet fermionic DM mass $m_{\psi}$. 
In the figure, the red points are the predictions for the scattering cross section of 
singlet fermionic DM and they are required to satisfy the measured DM mass density (\ref{wmap}). 
The cyan regions are consistent with the DAMA signals. Also denoted are the upper limits
from various other null experiments. 
We clearly see that the singlet fermionic DM can explain the DAMA signals 
without conflicting with other null experiments and 
simultaneously satisfy the DM mass density which is consistent
with the WMAP data.

\section{conclusion}
\label{sec:conclusion}

We have shown that a simple model of light singlet fermionic dark matter can 
reconcile the DAMA signal with other null experiments, while providing 
a right amount of DM mass density which is consistent with the WMAP observation.
The t-channel pair annihilation of the light dark matters 
to the pair of light singlet Higgs particles, 
with subsequent Higgs decays to the SM particles, 
would provide a large enough annihilation cross section 
to have a right amount of DM relic density. 
The light singlet fermionic DM, with the light singlet-like Higgs, 
can provide the spin-independent DM-nucleon scattering cross section 
which is compatible with the DAMA signal region and other null experiments, simultaneously.

\begin{acknowledgments}
This work was supported by the Korea Research Foundation Grant funded 
by the Korean Government (MOEHRD, Basic Research Promotion Fund) 
(KRF-2005-210-C000006 and KRF-2007-341-C00010), 
the Center for High Energy Physics of Kyungpook National University, 
and the BK21 program of Ministry of Education.
\end{acknowledgments}

\end{document}